\documentclass[a4paper]{article}
\usepackage[affil-it]{authblk}

\usepackage{graphicx}
\usepackage{graphics}
\usepackage{amsmath,amssymb}
\usepackage[usenames]{color}
\usepackage{subfigure}
\usepackage{hyperref}
\usepackage{enumitem}

\numberwithin{equation}{section}

\def\Tr{\,{\rm Tr}\,}

\newcommand{\be}{\begin{equation}}
\newcommand{\ee}{\end{equation}}
\newcommand{\bea}{\begin{eqnarray}}
\newcommand{\eea}{\end{eqnarray}}
\newcommand{\ben}{\begin{enumerate}}
\newcommand{\een}{\end{enumerate}}
\newcommand{\bit}{\begin{itemize}}
\newcommand{\eit}{\end{itemize}}

\newcommand{\la}[1]{\label{#1}}

\newcommand{\Eq}[1]{Eq.~(\ref{#1})}

\newcommand{\Sec}[1]{Sec.~\ref{#1}}

\newcommand{\Fig}[1]{Fig.~\ref{#1}}

\def\nn{\nonumber \\ }

\newcommand{\vv}[1]{\mathbf #1}						

\newcommand{\bert}{\raise-0.45mm\hbox{\Large$\Box$}}	

\definecolor{BrickRed}{cmyk}{0,0.89,0.94,0.28}
\definecolor{MidnightBlue}{cmyk}{0.98,0.13,0,0.43}
\definecolor{DarkGreen}{rgb}{0.100806,0.495968,0.209979}
\definecolor{orange}{rgb}{0.587167,0.354498,0.146197}

\begin{document}

\title{Interaction of a quantum field with a rotating heat bath}

\author{Robert Alicki\thanks{E-mail: \texttt{fizra@ug.edu.pl}}}

\affil{Institute of Theoretical Physics and Astrophysics, University of Gda\'nsk, 80-952 Gda\'nsk, Poland}

\author{Alejandro Jenkins\thanks{E-mail: \texttt{alejandro.jenkins@ucr.ac.cr}}}

\affil{Escuela de F\'isica, Universidad de Costa Rica, 11501-2060, San Jos\'e, Costa Rica\\
and Academia Nacional de Ciencias, 1367-2050, San Jos\'e, Costa Rica}

\date{\vspace{-5ex}}		

\maketitle


\begin{abstract}

The linear coupling of a rotating heat bath to a quantum field is studied in the framework of the Markovian master equation for the field's non-unitary time evolution.  The bath's rotation induces population inversion for the field's low-energy modes.  For bosons, this leads to superradiance, an irreversible process in which some of the bath's kinetic energy is extracted by spontaneous and stimulated emission.  We find the energy and entropy balance for such systems and apply our results to the theory of black hole radiation.  We also comment on how this relates to classical self-oscillations, including shear flow instabilities in hydrodynamics.

\end{abstract}

\section{Introduction}
\la{sec:intro}

Zel'dovich predicted in 1971 that a rotating black hole (BH) would radiate \cite{Zeldovich1,Zeldovich2}.  His reasoning was based on the observation that the same physics that causes damping of an incident electromagnetic field by a static dielectric implies that, if the dielectric's surface moves faster than the incident field's phase velocity, then the incident field will be amplified at the expense of the dielectric's kinetic energy.  Amplification of radiation by the sort of process that Zel'dovich described is often called ``superradiance'', a term introduced by Misner in 1972 \cite{Misner}.  It is an irreversible process, distinct from the equilibrium phenomenon, also identified as ``superradiance'', first described by Dicke in \cite{Dicke} and about which we will have nothing to say here.

Zel'dovich's thermodynamic argument was notably refined by Bekenstein and Schiffer in \cite{Bekenstein}.  More recently, the irreversibility of electromagnetically superradiant systems has been carefully investigated in \cite{Maghrebi1, Maghrebi2}.  For a thorough, modern review of rotational superradiance and its applications (with an emphasis on gravitational physics), see \cite{BCP}.

Zel'dovich's argument applies to any object capable of damping a classical or bosonic degree of freedom.  Some instances are Cherenkov radiation \cite{Cherenkov, QCherenkov}, sonic booms and Mach shocks \cite{shocks}, the Landau criterion for superflows \cite{Landau}, the Moon's tidal acceleration \cite{moon}, the shear flow instability by which the wind makes waves on the ocean's surface (see \Sec{sec:flow} in this paper), and mechanical instabilities of rotors.\footnote{One of the authors (AJ) thanks his student Carlos D\'iaz for help understanding the treatment of this particular subject in the mechanical engineering literature, where the analogy with superradiance has not been noted. A thermodynamic approach may help to generalize and simplify such analyses. \cite{DSTA}}  The prediction of BH superradiance motivated Hawking's subsequent discovery that a static BH must radiate thermally \cite{Hawking}.\footnote{For first-hand historical accounts of this, see \cite{BHT,Kip-book}.}

The novelty of our approach here is to provide a general and self-contained treatment of superradiance, based on the linear coupling of a quantum field to a rotating heat bath.  Our computations rely on the formalism of the Markovian master equation (also called the ``Lindblad equation'', after one of its developers) for an open quantum system \cite{Davies, Lindblad, GKS}.  Our results help illuminate the necessary connection between superradiance and the thermal dynamics of the bath, which for a stationary BH lead to Hawking radiation.  We also clarify the connection between superradiance of bosonic quantum fields and classical self-oscillations, using as example a flow instability.  This will allow us to clarify an important point about which there is some confusion in the hydrodynamics literature.

\section{Model and approximations}
\la{sec:model}

Consider a quantum field, either bosonic or fermionic, interacting with a source that acts as an equilibrium heat bath.  Let the heat bath rotate with angular velocity $\Omega$ about its symmetry axis, which we take to be the $z$-axis.  The free quantum field is described by the set of annihilation and creation operators $a_{m \alpha}(\omega)$, $ a^\dagger_{m \alpha}(\omega)$, corresponding to the field modes $| \omega, m, \alpha \rangle$ and satisfying the (anti-)commutation relations
\bea
\left[ a_{m \alpha}(\omega),  a^\dagger_{m' \alpha'}(\omega') \right]_{\pm} &=& \delta_{\omega \omega'} \delta_{m m'} \delta_{\alpha \alpha'}  , \nn
\left[ a_{m \alpha}(\omega),  a_{m' \alpha'}(\omega') \right]_{\pm} &=& [a^\dagger_{m \alpha}(\omega),  a^\dagger_{m' \alpha'}(\omega')]_{\pm} = 0 .
\la{eq:accr1}
\eea
We have set $\hbar = 1$ and written
\bea
\left[A, B \right]_+ &\equiv& [A, B] \equiv AB - BA \nn
\left[A, B \right]_- &\equiv& \{A, B\} \equiv AB + BA,
\la{eq:acomm}
\eea
for the commutator and anti-commutator respectively, so that in \Eq{eq:accr1} the sign $(+)$ corresponds to bosons and $(-)$ to fermions.  The quantum number $m = 0, \pm 1 , \pm 2 , ...$ indicates the angular momentum along the $z$-axis (which we take to be an axis of symmetry), $\omega \ge 0 $ the energy, and $\alpha$ the spin together with any other quantum numbers needed to specify the field's state.

For the free field, the Hamiltonian is
\be
H_{\rm f} = \sum_{\omega, m,\alpha} \omega\, a^\dagger_{m\alpha}(\omega) a_{m\alpha}(\omega)
\la{eq:hamF}
\ee
and the $z$-component of the angular momentum is
\be
L_{\rm f}^z =\sum_{\omega, m,\alpha} m\, a^\dagger_{m\alpha}(\omega) a_{m\alpha}(\omega) .
\la{eq:Lzf}
\ee
The most general field-bath interaction that is linear in the field operators is given by an additional term in the Hamiltonian of the form
\be
H_{\rm int} = \sum_{\omega, m,\alpha} \left( a_{m\alpha}(\omega)\otimes B^\dagger_{m \alpha}(\omega) + a^\dagger_{m \alpha}(\omega) \otimes B_{m \alpha}(\omega) \right),
\la{eq:hamint}
\ee
where $B_{m \alpha}(\omega)$ is a suitable bath operator. The bath has its own Hamiltonian $H_{\rm b}$ and its own $z$-component of the angular momentum $L_{\rm b}^z$.  The linearity of \Eq{eq:hamint} with respect to the quantum field is a valid approximation as long as the field is sufficiently weak that its self-coupling (either induced by the medium's polarizability or direct as in gravity and non-Abelian gauge fields) can be neglected.  All our computations will be in this weak-field regime, though we shall return to the issue of non-linearity in \Sec{sec:flow}.

Assuming that the interaction described by \Eq{eq:hamint} is invariant under rotations with respect to the $z$-axis, the bath operators may be chosen so that
\be
\left[ L^z_{\rm b} , B_{m \alpha}(\omega) \right] = -m B_{m \alpha}(\omega), \quad \left[ L^z_{\rm b} , B^\dagger_{m \alpha}(\omega) \right] = m B^\dagger_{m \alpha}(\omega),
\la{eq:BLcomm}
\ee 
and therefore
\bea
e^{i\phi L_{\rm b}^z} B_{m \alpha}(\omega) e^{-i\phi L_{\rm b}^z} &=& e^{i m\phi} B_{m \alpha}(\omega), \nonumber \\
e^{i\phi L_{\rm b}^z} B^\dagger_{m \alpha}(\omega) e^{-i\phi L_{\rm b}^z} &=& e^{-i m\phi} B^\dagger_{m \alpha}(\omega).
\la{eq:eigen}
\eea
To take into account the bath's rotation we use an effective Hamiltonian of the form
\be
H^{\rm eff}_{\rm b} = H_{\rm b} - \Omega L_{\rm b}^z,
\la{eq:Heff}
\ee
where the internal $H_{\rm b}$ is independent of the rotation or, at most, parametrically dependent on $\Omega$.  The arbitrary sign in front of the $\Omega$ in \Eq{eq:Heff} is taken negative for later notational convenience.

Equation \eqref{eq:Heff} is a valid approximation as long as the bath's coherent rotation does not involve energies sufficiently large to excite internal degrees of freedom.  This is analogous to the Born-Oppenheimer approximation in molecular physics, in which the electronic degrees of freedom are taken to be decoupled from the molecule's rotational and vibrational degrees of freedom \cite{Weinberg}.  We show in \Sec{sec:balance} that the kinetic energy of rotation can be converted into internal heating of the bath, but in our approximation this happens only via the coupling to the quantum field's low-energy modes.

Note that the Hamiltonian for our model is not necessarily quadratic, because $B_{m \alpha}(\omega)$ in \Eq{eq:hamint} is arbitrary.  The field's Markovian master equation that we derive for the weak-field limit in \Sec{sec:MME} applies, therefore, to models that are not exactly solvable.  Even for exactly solvable models, our techniques help us to understand and to characterize the relevant thermodynamics.

\section{Coupling spectrum and KMS condition}
\la{sec:coupling}

In the weakly coupled regime, all of the physically relevant properties of the bath are encoded in the second-order correlations of the bath operators $B_{m \alpha}(\omega)$, $B^\dagger_{m \alpha}(\omega)$, evaluated for the bath's stationary state \cite{Davies, AL2007}.  Using a short-hand notation $B_k , B^\dagger_l$ with multi-index $k \equiv \{\omega, m, \alpha \}$, these correlations are expressed as matrix elements of the coupling spectrum
\be
\gamma^0_{kl}(x) = \int_{-\infty}^{\infty} e^{ixt} \left\langle B^\dagger_k (t) B_l \right\rangle_{\rm b} dt ,
\la{eq:cspectrum}
\ee
were $\langle \ldots \rangle_{\rm b}$ denotes the expectation value with respect to the given stationary state of the bath, while $B_k(t)$ is the bath observable $B_k$ as it evolves in the Heisenberg picture for a bath governed by its internal Hamiltonian $H_{\rm b}$.  The matrix element $\gamma^0_{kl}(x)$ describes dissipative effects for frequency $x$, so that \Eq{eq:cspectrum} is an instance of the fluctuation-dissipation theorem \cite{Kubo}.

For the non-rotating bath in an equilibrium state, the following Kubo-Martin-Schwinger (KMS) condition is satisfied
\be
\gamma^0_{kl}(-x) = e^{-\beta x}\gamma^0_{-l-k}(x), \quad \mathrm{for} ~ B_{-k} \equiv B_k^\dagger,
\la{eq:KMS}
\ee
where $\beta$ is the inverse temperature of the stationary bath and $-k \equiv \{\omega, -m, \mathsf T \alpha \}$, with $\mathsf T \alpha$ the time-reversal of $\alpha$. Equation \eqref{eq:KMS} relates the rates of decay and its time-reverse (i.e., pumping) for a field coupled to the stationary bath. \cite{AL2007, BR}

The coupling spectrum matrix is diagonal in $m$ because of rotational symmetry.  The weak coupling approach eliminates from the Markovian master equation (discussed in the next section) the non-diagonal elements in $\omega$ (the ``secular approximation''). Finally, the non-diagonal elements in $\alpha$ can be removed by a suitable choice of the modes.  For simplicity, we use the same notation for this diagonal basis as for the original modes.

Replacing in \Eq{eq:cspectrum} the internal Hamiltonian by the effective one of \Eq{eq:Heff} and applying rotational invariance (see \Eq{eq:eigen}) we obtain a modified diagonal coupling spectrum for the rotating bath
\be
\gamma^\Omega_{kk}(x) = \gamma^0_{kk} (x+m\Omega).
\la{eq:cspectrum1}
\ee
This coupling spectrum satisfies a modification of the KMS condition of \Eq{eq:KMS}, namely
\be
\gamma^{\Omega}_{kk}(-x) = e^{-\beta (x-m\Omega)}\gamma^{\Omega}_{-k-k}(x).
\la{eq:KMS1}
\ee
%

\section{Markovian master equation}
\la{sec:MME}

Using Davies's weak coupling limit technique \cite{Davies}, one may derive the Markovian master equation (MME) for the density matrix $\rho(t)$ that describes the quantum field and acts on the corresponding Fock space.  For the reader unfamiliar with this formalism, more details on the derivation and interpretation of the MME are provided in the appendix.

For the model in \Sec{sec:model} we obtain that
\bea
&\dot \rho(t)& \hskip -8 pt = -i \left[ H_{\rm f}, \rho(t) \right] + \mathcal L \rho(t) = -i \left[ H_{\rm f}, \rho(t) \right] \nonumber \\
&& + \frac 1 2 \sum_{m, \alpha, \omega} \Gamma^{\Omega}_{m \alpha}(\omega)
\Bigl\{ \left( \left[ a_{m \alpha}(\omega) \rho(t), a^\dagger_{m \alpha}(\omega) \right] + \left[ a_{m \alpha}(\omega), \rho(t)a^\dagger_{m \alpha}(\omega) \right] \right) \nonumber \\
&& + e^{-\beta(\omega - m\Omega)} \left( \left[ a^\dagger_{m \alpha}(\omega) \rho(t), a_{m \alpha}(\omega) \right] + \left[ a^\dagger_{m \alpha}(\omega), \rho(t)a_{m \alpha}(\omega) \right] \right) \Bigl\} .
\la{eq:MME}
\eea
Here $\mathcal L$ is short-hand notation for the dissipative part of the Lindblad superoperator, which induces non-unitarity in the time evolution of the quantum state described by $\rho(t)$ \cite{Lindblad,GKS}.  For convenience, we have used the same symbol $H_{\rm f}$ for the renormalized Hamiltonian in \Eq{eq:MME}, which contains corrections due to the field-bath interaction, as for the bare Hamiltonian in \Eq{eq:hamF}.  The symbol $\Gamma^{\Omega}_{m \alpha}(\omega)$ in \Eq{eq:MME} denotes the diagonal element of the coupling spectrum matrix of \Eq{eq:cspectrum1} with $k = \{ \omega, m ,\alpha \}$, evaluated at frequency $x = \omega$.
\par
The quantity
\be
\gamma_\downarrow (k) \equiv \Gamma^{\Omega}_{m \alpha}(\omega) = \gamma_{k k}^0 ( \omega + m \Omega)
\la{eq:decay}
\ee
is the decay rate for the mode $k$, whereas
\be
\gamma_\uparrow (k) \equiv e^{-\beta (\omega - m \Omega)} \Gamma^{\Omega}_{m \alpha}(\omega) = e^{-\beta (\omega - m \Omega)} \gamma_\downarrow (k)
\la{eq:pumping}
\ee
is the corresponding pumping rate.  The ratio of the pumping to the damping rates in \Eq{eq:MME} may be expressed as a Boltzmann factor with ``local'' (i.e., $\omega$-dependent) inverse temperature $\beta_{\rm loc}[\omega]$, such that
\be
e^{-\beta_{\rm loc}[\omega] \omega} \equiv \frac{\gamma_\uparrow (k)}{\gamma_\downarrow (k)} = e^{-\beta(\omega - m\Omega)} ,
\la{eq:Boltzmann}
\ee
implying that
\be
\beta_{\rm loc}[\omega] = \beta \left(1- \frac{m\Omega}{\omega} \right).
\la{eq:Boltzmann1}
\ee
Thus,
\be
\beta_{\rm loc}[\omega] < 0 ~~ \Leftrightarrow ~~ \omega < m\Omega .
\la{eq:superradiant}
\ee
Such a negative local temperature for the rotating bath indicates a population inversion in the low-energy modes with $\omega < m \Omega$.  We shall see in \Sec{sec:stability} that this leads to a superradiant instability for bosonic fields (in which the Hamiltonian of a single mode is unbounded).  In quantum optics it is common to describe the population inversion of a lasing medium in terms of its negative local temperature (see \cite{localT} for a general theory).

The modes with $\beta_{\rm loc}[\omega] < 0$ are those for which the energy $\omega'$ measured in the bath's comoving frame of reference is negative \cite{Zeldovich1}.  It is not surprising, therefore, that their equilibrium populations should be inverted, but our treatment clarifies how this depends on the field-bath interaction: note that we had to assume rotational invariance of the interaction in \Eq{eq:hamint}, and a large energy-scale separation between the bath's rotational and internal degrees of freedom (\Eq{eq:Heff}).  We shall see that a thermodynamically complete description of superradiance must incorporate a {\it positive feedback} between field and bath, which in the quantum picture is provided by stimulated emission.

\section{Stable and unstable modes}
\la{sec:stability}

The average occupation number of a single mode
\be
\bar n_{m \alpha}(\omega, t) \equiv \Tr \left[ \rho(t) a^\dagger_{m \alpha}(\omega) a_{m \alpha}(\omega) \right] 
\la{eq:pnumber}
\ee
obeys the equation
\bea
\dot{\bar n}_{m \alpha}(\omega, t) &=& 
- \left( \Gamma^{\Omega}_{m \alpha}(\omega) \left[ 1 - (\pm) e^{-\beta(\omega - m\Omega)} \right] \right) \bar n_{m \alpha}(\omega, t) \nonumber \\
&& + \Gamma^{\Omega}_{m \alpha}(\omega)e^{-\beta(\omega - m\Omega)},
\la{eq:pnumber1}
\eea
where, as in Eqs.\ \eqref{eq:accr1} and \eqref{eq:acomm}, the sign $(+)$ corresponds to bosons and $(-)$ to fermions.

The fermionic solution to \Eq{eq:pnumber1} for $ t\geq 0$ is
\bea
\bar n_{m \alpha}(\omega, t) &=&
\exp \left\{ -\Gamma^{\Omega}_{m \alpha}(\omega) \left[ 1 + e^{-\beta(\omega - m\Omega)} \right] t \right\} \bar n_{m \alpha}(\omega, 0) \nonumber \\
&+& \left[ 1- \exp \left\{ -\Gamma^{\Omega}_{m \alpha}(\omega) \left[ 1 + e^{-\beta(\omega - m\Omega)} \right] t \right\} \right] \bar n_{m \alpha}(\omega, \infty),
\la{eq:fermions}
\eea
with asymptotic population
\be
\bar n_{m \alpha}(\omega, \infty) = \left[ e^{\beta(\omega - m\Omega)} + 1 \right]^{-1}
\la{eq:fermions1}
\ee
corresponding to the Fermi-Dirac distribution with inverse temperature $\beta$ and chemical potential $m \Omega$. 

For bosons, on the other hand, only modes $| \omega, m, \alpha \rangle$ that satisfy
\be
\omega > m \Omega
\la{eq:stable}
\ee
are stable.  In that case the solution is
\bea
\bar n_{m \alpha}(\omega, t) &=&
\exp \left\{ -\Gamma^{\Omega}_{m \alpha}(\omega) \left[ 1 - e^{-\beta(\omega - m\Omega)} \right] t \right\} \bar n_{m \alpha}(\omega, 0) \nonumber \\
&+& \left[ 1- \exp \left\{ -\Gamma^{\Omega}_{m \alpha}(\omega) \left[ 1 - e^{-\beta(\omega - m\Omega)} \right] t \right\} \right] \bar n_{m \alpha}(\omega, \infty),
\la{eq:bosons}
\eea
with asymptotic population
\be
\bar n_{m \alpha}(\omega, \infty) = \left[ e^{\beta(\omega - m\Omega)} -1 \right]^{-1} ,
\la{eq:bosons1}
\ee
corresponding to the Bose-Einstein distribution with inverse temperature $\beta$ and chemical potential $m\Omega$.  Modes satisfying the condition
\be
\omega < m\Omega ,
\la{eq:unstable}
\ee
on the other hand, are unstable and their occupation numbers grow exponentially with time.  This corresponds to Zel'dovich's rotational superradiance.

It is interesting to consider the case of zero temperature, i.e., the limit $\beta \to \infty$ in \Eq{eq:pnumber1} for superradiant modes.  Using the KMS conditions of Eqs.\ (\ref{eq:KMS}) and (\ref{eq:KMS1}), one obtains for the pumping rate
\be
\gamma_{\uparrow} (k) = \Gamma^{\Omega}_{m \alpha}(\omega) e^{-\beta(\omega - m\Omega)} = \gamma^0_{-k-k}(m\Omega -\omega) \equiv \gamma_{\omega m \alpha} (m \Omega - \omega) ,
\la{eq:zerolimit}
\ee
where $\gamma_{\omega m \alpha}(x)$, for $x \geq 0$ is the damping rate of the mode $| \omega, -m, \mathsf T\alpha \rangle$ at the frequency $x$.  Note that this damping rate remains positive in the zero-temperature limit, while $\gamma_{\omega m \alpha}(-x)$ tends to zero.  Thus, the rate in \Eq{eq:pnumber1}, in the zero-temperature limit, becomes
\be
\dot{\bar n}_{m \alpha}(\omega, t) =   \gamma_{\omega m \alpha} (m\Omega -\omega) \left[1 + \bar n_{m \alpha}(\omega, t) \right] .
\la{eq:pnumber2}
\ee
Equation \eqref{eq:pnumber2} implies that a rotating body at zero temperature will produce a continuous spectrum of radiation, with a non-trivial spatial distribution determined by the superradiant condition of \Eq{eq:unstable}.

Our results are consistent with those obtained in \cite{Endlich} by the methods of effective quantum field theory.  There the authors conclude that the probability of absorption by an object at rest entirely determines the superradiant amplification when the same object is rotating.  We believe that our own formulation of superradiance in terms of the MME offers a more transparent thermodynamic interpretation and may therefore be more readily generalizable, including to non-relativistic and to classical systems.

\section{Feedback and stimulated emission}
\la{sec:feedback}

The structure of \Eq{eq:MME} implies that the different field modes $|\omega, m, \alpha \rangle$ evolve independently. Moreover, the diagonal matrix elements of $\rho(t)$, computed in the corresponding population number representation basis, give the probabilities $P_n(k;t)$ of finding $n$ particles in a given mode $|k \rangle \equiv| \omega, m, \alpha \rangle$.  These probabilities evolve according to a Markovian birth-death process
\bea
&\dot P_n(k;t)& \hskip -8 pt = \gamma_\downarrow (k) (n +1) P_{n+1}(k;t) + \gamma_\uparrow (k) \left[ 1\pm(n-1) \right] P_{n-1}(k;t) \nonumber \\
&& - \left[ \gamma_\downarrow (k) n +\gamma_\uparrow (k) (1\pm n) \right] P_{n}(k;t).
\la{eq:birth}
\eea
In the symbol $\pm$ on the right-hand side of \Eq{eq:birth}, the sign $(+)$ corresponds to bosons (with unbounded population \hbox{$n=0,1,2,\ldots$}) and the sign $(-)$ to fermions (with bounded population $n=0,1$).

From the last term on the right-hand side of \Eq{eq:birth} we see that the probability, per unit time, of creating a new particle if $n$ particles are already present in a given mode depends on $n$ and is equal to $\gamma_{\uparrow}(k)(1\pm n)$, where the 1 corresponds to spontaneous emission.  This $n$-dependence can be interpreted as resulting from a feedback between the field and the bath.  This feedback is positive for bosons, due to stimulated emission.  For fermions the feedback is negative, due to the Pauli exclusion principle.

Superradiance is a process in which the kinetic energy of the heat bath's rotation powers coherent radiation modes, without requiring any resonant tuning.  In a classical context, such processes are described as ``self-oscillations''.  For a review of self-oscillation as a non-equilibrium process depending on a positive feedback between the oscillator and the power source, see \cite{SO}.

\section{Energy and entropy balance}
\la{sec:balance}

The formal Gibbs state 
\be
\bar{\rho} = Z^{-1} e^{-\beta \left( H_{\rm f} - \Omega L_{\rm f}^z \right)}
\la{eq:gibbs}
\ee
is a stationary solution of \Eq{eq:MME}. This allows us to apply the general formula for the entropy production based of the Spohn inequality \cite{Spohn, Alicki},
\be
\sigma(t) = - \Tr \left[ \mathcal L \rho(t) \left( \ln \rho(t) - \ln \bar{\rho} \right) \right] \geq 0 ,
\la{eq:spohn}
\ee
in order to derive the entropy balance (i.e., the second law of thermodynamics).  In units such that $k_B = 1$, this gives
\be
\dot S(t) = \sigma(t) + \beta J , 
\la{eq:IIlaw}
\ee
where $S(t) \equiv -\Tr \left[ \rho(t) \ln \rho(t) \right]$ is the entropy of the field, while
\be
J = \frac{d}{dt} \Tr \left[ \rho(t) \left(H_{\rm f} - \Omega L_{\rm f}^z \right) \right] .
\la{eq:heat}
\ee
is the heat current flowing from the bath and into the field.  Identifying the internal energy $U$ of the field with its average value of the Hamiltonian,
\be
U(t) =  \Tr \left[ \rho(t) H_{\rm f} \right] , 
\la{eq:U}
\ee
we obtain the energy balance (i.e., the first law of thermodynamics)
\be
\dot U(t) =  J + \Omega \frac{d}{dt} \Tr \left[ \rho(t) L_{\rm f}^z \right], 
\la{eq:Ilaw}
\ee
where the second term on the right-hand side of \Eq{eq:Ilaw} is the power supplied by the rotating bath.

Note that the Gibbs state of \Eq{eq:gibbs} cannot be normalized for bosonic modes with $\omega < m \Omega$ (i.e., for the superradiant modes). However, introducing intermediate cutoffs in particle numbers one can derive rigorously the results of Eqs.\ (\ref{eq:IIlaw}) and (\ref{eq:Ilaw}).

Combining \Eq{eq:heat} with the kinetic \Eq{eq:pnumber2} provides a simple derivation of the fact that, at zero temperature, superradiance is accompanied by a heating of the bath.  Indeed, the corresponding expression for the heat current into the field,
\be
J = \sum_{\{\omega< m \Omega, m, \alpha\}} (\omega - m\Omega) \cdot \gamma_{\omega m \alpha} (m\Omega -\omega) \cdot \left[1 + \bar n_{m \alpha}(\omega, t) \right],
\la{eq:heat0}
\ee
is evidently negative.

In comparing our results to superradiant spectra computed by other authors for specific systems, one should bear in mind that we formulated the model of \Sec{sec:model} in terms of the field modes as they would exist within a cavity containing the heat bath.  The non-unitarity represented by the superoperator $\cal L$ in \Eq{eq:MME} comes from the dynamical interaction with the bath, without any modification of the field's boundary conditions.  Other treatments often frame superradiance as a scattering problem, with at least some of the non-unitarity given by the choice of boundary conditions within the bath; see, e.g., \cite{Manogue, Unruh}.

\section{Black hole radiation}
\la{sec:BH}

For any quantum field placed around a BH, its modes can be separated into those localized far outside the BH (outer modes) and those close or below the event horizon (inner modes). The particles occupying the inner modes form a bath for the outer ones. The interaction between both systems can be treated as a kind of tunneling process described by the quadratic Hamiltonian of the form of \Eq{eq:hamint}, with the operator $a^\dagger_{m \alpha}(\omega)$ creating a particle in an outer mode and $B_{m \alpha}(\omega)$ annihilating a particle in a certain superposition of inner modes.

As shown by Hawking in \cite{Hawking}, the strong gravity of a BH creates indeterminacy between the  annihilation operator $b_k$ and the creation operator $b^\dagger_k$ of an inner mode with multi-index $k$. Hence, we can introduce a bath operator in \Eq{eq:hamint} of the form
\be
B_k =\sum_{k'=\{\omega', m ,\alpha'\}} \left[ f_{k}(\omega') b_{k'} + g_{k}(\omega') b^\dagger_{-k'} \right]
\la{eq:BHop}
\ee
with some form factors $f_{k}(\omega), g_{k}(\omega)$ (as before, $-k$ stands for the time-reversal of the quantum numbers in $k$).  There is no summation over $m' \neq m$ in \Eq{eq:BHop} because we assumed rotational symmetry.

The key result of \cite{Hawking} is that the indeterminacy between the annihilation and creation operators is exponentially small and approximately given by
\be
\frac{|g_{k}(\omega)|^2}{|f_{k}(\omega)|^2} = e^{-\beta_{\rm H} \omega}
\la{eq:BHtemp}
\ee
where $\beta_{\rm H}$ is the inverse Hawking temperature of the BH. Inserting Eqs.~\eqref{eq:BHop} and \eqref{eq:BHtemp} into \Eq{eq:cspectrum} for the vacuum state $| 0 \rangle$ of the inner modes (i.e. the state $| 0 \rangle$ such that $b_k | 0 \rangle = 0$ for all $k$'s), one obtains the KMS condition of \Eq{eq:KMS} with $\beta = \beta_{\rm H}$, thereby establishing that the static black hole behaves with respect to the external fields as a heat bath at the Hawking temperature.

The results of \Sec{sec:stability} then imply that a Kerr BH will superradiate bosons obeying the condition of \Eq{eq:unstable}, with a spectrum determined by the form factors in \Eq{eq:BHop}.  The negative energy of a superradiant mode as measured in the Kerr BH's co-moving frame (see \Sec{sec:MME}) reflects the presence of an ergosphere allowing extraction of the BH's angular momentum by the Penrose process \cite{Penrose}.

In the thermodynamic relation for a Kerr-Newman BH with mass $M$, angular momentum $L$, charge $Q$, event horizon area $A$, and Hawking temperature $T_{\rm H}$,
\be
dM = T_{\rm H} \frac{dA}{4} + \Omega dL + \Phi dQ ,
\la{eq:BH-thermo}
\ee
the angular velocity $\Omega$ and the electrostatic potential $\Phi$ near the event horizon appear as chemical potentials.  (In \Eq{eq:BH-thermo} the fundamental constants $c$, $\hbar$, and $G$ have all been set to 1.)  Rotational and charged superradiance are processes that extract some of the BH's internal energy ($dM < 0$), while  increasing the entropy within the BH's event horizon ($dA > 0$) \cite{charged-super}; see also \cite{BCP} and references therein.  The energy source is non-thermal, but the irreversibility of superradiance relates it to the purely thermal Hawking radiation of a BH with $\Omega = 0$ and $\Phi = 0$.  Since Hawking radiation does not depend on stimulated emission, it produces fermions as well as bosons.\footnote{The absence of fermionic superradiance (about which we will have more to say in \Sec{sec:flow}) appears, in the context of Hawking's area theorem \cite{area}, as tied to the fact that fermions violate the weak energy condition \cite{charged-super, neutrinos}.}

The relation between BH superradiance and Hawking radiation is analogous to the relation between the coherent radiation produced by a laser and the thermal radiation that the unpumped lasing medium at non-zero temperature would emit into a surrounding vacuum (a radiation associated with the non-zero $\gamma_\uparrow (k)$'s).  Though distinct, the two processes are tied by thermodynamic irreversibility.

\section{Classical limit and flow instabilities}
\la{sec:flow}

The shear flow instability that explains how the wind makes waves on the surface of the ocean is an interesting classical analog of the superradiance of bosonic fields.\footnote{John McGreevy brought this to our attention some years ago.}  It is instructive to consider how the thermodynamic arguments of \cite{Zeldovich2, Bekenstein, Maghrebi2} apply to this instability and to identify the hydrodynamical positive feedback, as well as the reasons why it leads to a result similar to that given by stimulated emission of bosons.

\begin{figure} [t]
\begin{center}
	\includegraphics[width=0.8 \textwidth]{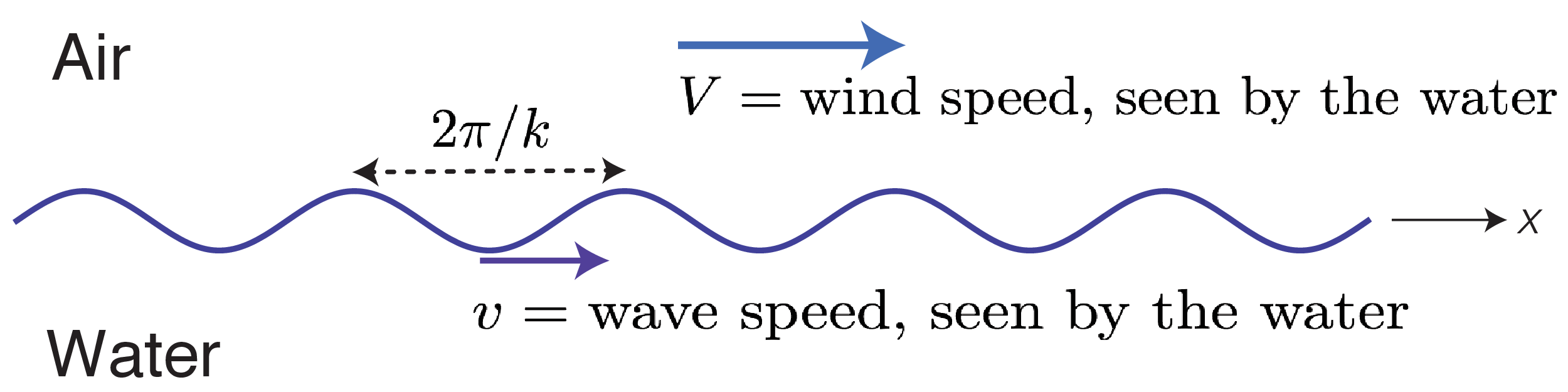}
\end{center}
\caption{\small Illustration of the shear flow instability by which wind can generate waves on the surface of a body of water.  Image adapted from \cite{Kip-talk}.\la{fig:shear}}
\end{figure}

The relevant instability can occur at the interface between two layers of fluid, as long as there is a difference in their respective tangential velocities.  Let us consider an upper layer of air with mass $m$, which moves with horizontal velocity $V$ with respect to the water below, as shown in \Fig{fig:shear}.  Non-relativistically, the air's momentum is $\vv p_{\rm wind} = m \vv V$ and its corresponding kinetic energy is
\be
E_{\rm wind} = \frac{p^2}{2m}.
\la{eq:windE}
\ee
The air's viscosity provides a coupling between the air and the surface of the water.  Consider a traveling wave on the water's surface, which in the linear regime may be described by a complex-valued
\be
\xi \sim e^{i(kx - \omega t)},
\la{eq:travel}
\ee
where $x$ is the spatial coordinate in the horizontal direction, $k$ the wavenumber, and
\be
v = \frac \omega k
\la{eq:phasev}
\ee
the phase velocity.  Translational invariance implies momentum conservation
\be
\dot {\vv p}_{\rm wind} = - \dot {\vv p}_{\rm wave}.
\la{eq:pcons}
\ee

We may express the rate at which the wave is gaining momentum as
\be
\dot {\vv p}_{\rm wave} = f \hbar \vv k ,
\la{eq:pquant}
\ee
where $\hbar \vv k$ is the momentum of a quantum of wave and $f$ is the rate at which such quanta are being produced (for clarity, in this section we write the factors of $\hbar$ explicitly).  Taking the time derivative of \Eq{eq:windE} and using Eqs.\ \eqref{eq:pcons} and \eqref{eq:pquant} gives
\be
- \dot E_{\rm wind} = \frac{\vv p}{m} \cdot \dot {\vv p}_{\rm wave} = {\vv V} \cdot ( f \hbar \vv k)~
\la{eq:windP}
\ee
Since each wave quantum carries energy $\hbar \omega$, we also have that
\be
\dot E_{\rm wave} = f \hbar \omega = f \hbar v k
\la{eq:waveP}
\ee
Combining Eqs.\ \eqref{eq:windP} and \eqref{eq:waveP} we conclude that
\be
\left| \dot E_{\rm wind} \right| > \dot E_{\rm wave} ~ \Leftrightarrow ~ V > v .
\la{eq:diss}
\ee

The instability condition is
\be
V > v .
\la{eq:critic}
\ee
From \Eq{eq:diss} we can see that this is because, in such a case, dissipative processes can, while conserving momentum, extract more kinetic energy from one fluid layer (in our case, from the wind) than goes into amplifying waves on the interface.  That energy difference is available to generate entropy (here by viscous dissipation in the air) and the process therefore proceeds irreversibly.

This thermodynamic analysis is trivially generalizable to systems with a discontinuity in rotational rather than translational velocity and to relativistic systems, as in \cite{Zeldovich2, Bekenstein, Maghrebi2}.  Note that for a cylindrically symmetrical system with a linear dispersion relation for the waves, \Eq{eq:unstable} corresponds exactly to \Eq{eq:critic}, with $V$ the linear speed at which the surface of the inner cylinder moves with respect to the outer cylinder, and $v$ the phase speed of the waves along the azimuthal direction.  (Zel'dovich explicitly invoked such a correspondence in \cite{Zeldovich1}.)

Zel'dovich noted that quantization makes the argument particularly simple and concluded that this was an instance of how ``quantum mechanics helps understand classical mechanics'' \cite{Zeldovich1}.\footnote{He had previously published a brief, pseudonymous piece with that title \cite{Paradoksov}.}  Only the bosonic case, with no upper bound on the number of quanta with wavenumber $k$, can approach a classical regime.  It is interesting to consider how the feedback from stimulated emission, described in \Sec{sec:feedback}, appears in the classical flow instability.

Classical self-oscillation is due to a backreacting force that {\it leads} the oscillation and therefore amplifies it \cite{SO}.  In the shear flow instability of \Fig{fig:shear} this results from the bath moving faster than the wave's phase, which is equivalent to negativity of the energy
\be
\omega' = \frac i \xi \frac{\partial \xi}{\partial t'} = \frac{i}{\xi} \left( \frac{\partial \xi}{\partial t} - V \frac{\partial \xi}{\partial x} \right) = \omega \left( 1 - \frac V v \right)
\la{eq:omegap}
\ee
in the bath's comoving frame (as it does not affect our reasoning, we omit the relativistic factor $(1 - V^2/c^2)^{-1/2}$ in \Eq{eq:omegap}).  If $V > v$, then $\omega ' < 0$ and the oscillation sees the time lag resulting from intra-bath dissipation as a lead, thus establishing a positive feedback between the wave and the air pressure above it.

In the equation of motion for a scalar $\xi$, the condition $\omega' < 0$ also implies that the sign of the linear damping term $a (\partial \xi / \partial t)$ flips upon transformation to the bath's comoving frame (this was Zel'dovich's first argument for superradiance in \cite{Zeldovich1}).  A fermionic field's damping appears in its equation of motion as an imaginary part of the invariant mass, which therefore does not change sign for $\omega' < 0$.  Absence of fermionic superradiance may also be seen from the solutions to scattering off a potential barrier \cite{Manogue} or, in an AdS/CFT context, from the behavior of the poles of the retarded Green functions corresponding to boundary-theory operators \cite{McGreevy}.

In classical hydrodynamics, one may see the instability of \Fig{fig:shear} as a consequence of the fact that, when $v < V$, the air ``trapped'' in a valley of the wave is slowed down, causing the air pressure there to increase (by Bernoulli's theorem) relative to the crests.  This, combined with the lag of that air pressure oscillation with respect to the phase of the wave ---a lag resulting from the air's viscosity and therefore associated with dissipation in the air--- implies that the air pressure does net positive work on the wave over a full period.

Mollo-Christensen emphasized in \cite{Mollo-Christensen} that an inviscid wind can do no net work over a full period of the water's motion.  Textbook treatments find an instability for inviscid shear flow (the ``Kelvin-Helmholtz instability'') that in the center-of-momentum frame corresponds to a non-oscillatory divergence, and which predicts an unrealistically large critical wind speed for raising waves on the surface of body of water \cite{KH-TB}.  Long ago, Lamb pointed out that allowing for viscous dissipation in the air fixes the problem \cite{Lamb}, but without sufficient emphasis and clarity for the point to have been widely grasped.\footnote{It is well known in mechanical engineering that a system may be stable when dissipation is exactly zero and unstable for vanishingly small but positive dissipation.  For more on such ``dissipation-induced instabilities'' and their connection to superradiance, see \cite{DSTA}.}  The approach that we have advocated here may therefore help to clarify certain obscurities and confusions that persist in the literature on the roles of viscosity and turbulence in shear flow instabilities.

Linear feedback causes the wave's amplitude to grow exponentially with time.  Non-linear dissipative effects eventually limit the wave's growth, giving a steady amplitude that corresponds to a limit cycle in the classical phase space \cite{SO}.  The non-linear approach to a limit cycle is evidently an irreversible process that erases information about initial conditions and transient noise.  The initial, linear runaway is also irreversible, but to see that we must take into account the dynamics of the bath (the wind in \Fig{fig:shear}) in which entropy is generated.

To better understand the quantum-classical correspondence for such systems, consider the bosonic birth-death process described by \Eq{eq:birth}, but with the probability of single-boson absorption per unit time replaced by a more general expression
\be
\gamma_\downarrow (k) n ~ \rightarrow ~ \gamma_\downarrow (k) \left( n + \kappa n^2 + \ldots \right), ~~\hbox{for}~~ \kappa \geq 0.
\la{eq:nonlin}
\ee
This corresponds to going beyond the linear coupling of \Eq{eq:hamint}, from which we derived the simplest MME (\Eq{eq:MME}).  A $\kappa > 0$ in \Eq{eq:nonlin} can, for instance, describe gain saturation in a laser \cite{Carmichael}.

Denote by $\langle \cdots \rangle$ the average of an observable over the probabilities $P_n(k;t)$. For $\kappa > 0$ there is no closed kinetic equation for the average population $\langle n \rangle$ of a given mode, because of the presence of the higher-order term $\langle n^2 \rangle$.  However, in the semi-classical limit the fluctuations of the population numbers may be neglected, so that one can replace $\langle n^2 \rangle$ by $\langle n \rangle^2$.  This gives a kinetic equation for $\langle n \rangle$ that can be interpreted classically, with the quantum stimulated emission transformed into a non-linear feedback that gives rise to a limit cycle.  This is an instance of non-linear classical mechanics emerging from the linear evolution of quantum states.

The general relation between the kinetic equations derived from the Markovian birth-death process for the quantum system and the hydrodynamical description of the system lies beyond the scope of this paper.  An intriguing conceptual question in this regard is how the quantum picture of feedback, based on stimulated emission, translates into a classical picture in terms of a back-reacting, phase-shifted force.  But the argument presented here illustrates clearly how the large-$n$ limit of bosonic superradiance is directly connected to a hydrodynamic instability, and why dissipation is required in a consistent hydrodynamic description.

\section{Discussion}
\la{sec:discussion}

Superradiance is by now a well-known effect in high-energy physics, general relativity, and quantum optics.  The contribution of this article was to derive it in the formalism of the MME for the quantum field coupled to a  heat bath with a rotational symmetry that admits an effective description of the form of \Eq{eq:Heff}.  Rather than assuming the laws of thermodynamics, we have derived them (see \Sec{sec:balance}) from the dynamics of the field as an open quantum system in the Markovian limit.  This clarified how superradiance depends on the feedback between the field and the bath, and on the generation of entropy in the bath, allowing us to clarify a point on which confusion persists even in the best textbook treatments of flow instabilities (see \Sec{sec:flow}).

This approach also allowed us to fix the possible dependence of the superradiant spectrum on the rotational velocity $\Omega$, which appears exclusively through the frequency shift of \Eq{eq:cspectrum1}.  In \Sec{sec:flow} we clarified the connection between superradiance as a quantum process that proceeds via stimulated emission, and classical self-oscillations such as flow instabilities.  Our results sharpen the arguments of \cite{BCP, Unruh} that superradiance requires not just the presence of negative-energy states (associated, in a Kerr BH geometry, to the ergosphere), but also non-unitarity in the field's time evolution (given, in a BH geometry, by the event horizon).  We saw that in the weak-field, linear-response approximation this non-unitarity, represented by the $\cal L$ in the MME of \Eq{eq:MME}, has a clear and consistent thermodynamic interpretation.  In \Sec{sec:flow} we commented on how non-linearity is related to gain saturation and to the approach to a classical limit cycle, but a detailed understanding of this non-linear regime remains a challenge to be met by future research.

The thermodynamic point of view allows us to abstract much of the detailed physics and to treat many distinct systems within a simple and unified framework.  The wisdom of this approach has been demonstrated in recent decades by the fruitfulness of BH thermodynamics \cite{BH-thermo}.\footnote{Thorne's account of Zel'dovich's original argument for BH superradiance in \cite{Kip-book} brings to mind an aphorism that one of us (AJ) received as a student from Gil Refael: ``Thermodynamics is saying something without knowing anything.''}  Incorporating into this thermodynamical picture a more sophisticated understanding of the evolution of open quantum systems may shed light on fundamental problems.

\vskip 10 pt {\bf Acknowledgements:} We thank Mohammad Maghrebi for helpful comments and feedback on this work, and Carlos D\'iaz for help with the hydrodynamical literature and with understanding the basic distinction between the textbook Kelvin-Helmholtz instability and Zel'dovich's dissipation-induced instability applied to the same shear flow.  We also thank Blai Garolera, Jos\'e Gracia-Bond\'ia, Bob Jaffe, Zohar Komargodski, Karl Landsteiner, John McGreevy, Surjeet Rajendran, and Joe V\'arilly for discussions.  AJ acknowledges the hospitality of the Instituto de F\'isica Te\'orica (IFT), of the Universidad Aut\'onoma de Madrid (UAM) and the Consejo Superior de Investigaciones Cient\'ificas (CSIC) of Spain, while some of this work was being completed.  That visit was supported by the European Union's Horizon 2020 research and innovation program under the Marie Sk{\l}odowska-Curie grant agreement no.\ 690575.

\appendix
\section{Markovian Master Equation}
\la{sec:appendix}

For the reader's convenience, we present here a brief derivation of the Markovian Master Equation (MME) describing the dynamics of an open quantum system weakly coupled to a heat bath.  We then briefly sketch the thermodynamic interpretation of the MME, based on the Spohn inequality.  For further details, see \cite{AL2007}.  Throughout this appendix we set $\hbar = 1$.  

\subsection{MME in the weak coupling regime}

Consider a system with a bare Hamiltonian $H^0$ and a bath with Hamiltonian $H_R$, coupled together by an interaction Hamiltonian of the form
\be
\lambda H_{\rm int} = \lambda S \otimes R ,
\la{eq:hamint0}
\ee
where $S$ and $R$ are, respectively, Hermitian operators for the system and for the reservoir, with the constant $\lambda$ giving the strength of the coupling.  We also assume that
\be
[ \rho_R , H_R ] = 0 ~~\hbox{and}~~ \Tr (\rho_R\, R) = 0 .
\ee
It is straightforward to generalize this to any interaction of the form
\be
\lambda H_{\rm int} = \lambda \sum_\alpha S_\alpha \otimes R_\alpha
\la{eq:hamint1}
\ee
or, in what can sometimes be a more natural parametrization,
\be
\lambda H_{\rm int} = \lambda \sum_\alpha \left( S_\alpha \otimes R_\alpha^\dagger + S_\alpha ^\dagger \otimes R_\alpha \right) ,
\la{eq:hamint2}
\ee
for non-hermitian $S_\alpha,  R_\alpha$.  From now on we use \Eq{eq:hamint0} for notational simplicity.

In the interaction picture, the reduced dynamics for the system only is given by the partial trace over the sub-space of the bath:
\be
\rho (t) = \Lambda (t,0) \rho \equiv \Tr_R \left[ U_\lambda (t,0) \rho \otimes \rho_R U_\lambda (t,0)^\dagger \right] ,
\la{eq:red_dyn}
\ee
where the unitary propagator in the interaction picture is given by the time-ordered exponential
\be
U_\lambda (t,0) = {\cal T} \exp \left\{ -i \lambda \int_0^t S(s) \otimes R(s) \,ds \right\}
\la{eq:prop_int}
\ee
where
\be
S(t) = e^{i H t} S e^{- i H t} ,\  R(t)= e^{i H_R t} R e^{- i H_R t}.
\la{eq:prop_int1}
\ee
Notice, that $S(t)$ is defined not in terms of the bare $H^0$, but rather of the renormalized $H$, which can be expressed as
\be
H = H^0 + \lambda^2 H_{1}^{\rm corr} + \ldots  
\la{eq:H_S}
\ee
The terms containing powers of $\lambda $ in \Eq{eq:H_S} are Lamb-shift corrections due to the interaction with the bath, and which cancel with the uncompensated term $H - H^0$ that should, in principle, be present in \Eq{eq:prop_int}.  The lowest order (Born) approximation with respect to the coupling $\lambda$ yields $H_1^{\rm corr}$.  To find the higher order terms one must go beyond the Born approximation.

A convenient tool, though one not used in the rigorous derivations of the MME, is the cumulant expansion for the reduced dynamics
\be
\Lambda (t,0) = \exp \sum_{n=1}^\infty \left[ \lambda^n K^{(n)} (t) \right] .
\ee
One finds that $K^{(1)} = 0$.  The Born approximation (corresponding to weak coupling) consists of truncating the cumulant expansion at $n = 2$.  We denote $K \equiv K^{(2)}$, so that 
\be
\Lambda (t,0) = \exp \left[\lambda^2 K(t) + O \left( \lambda^3 \right) \right] .
\ee
One obtains
\be
K(t) \rho = \int_0^t ds \int_0^t du F(s-u) S(s) \rho S^\dagger (u) +(\rm{similar\ terms}) ,  
\la{eq:K(t)}
\ee
where $F(s)= \Tr [\rho _R R(s) R]$.  The ``similar terms'' in \Eq{eq:K(t)}) are of the form $\rho S(s) S^\dagger (u)$ and $S(s) S^\dagger (u) \rho$.

The Markovian approximation (in the interaction picture) consists of taking, for sufficiently long time $t$,
\be
K(t) \simeq t {\cal L} ,
\la{eq:L}
\ee
where $\cal L$ is the Lindblad-Gorini-Kossakowski-Sudarshan (LGKS) generator. To find its form we first decompose $S(t)$ into its Fourier components
\be
S(t) = \sum_{\{\omega\} } e^{i\omega t} S_\omega , ~~\hbox{with}~~  S_{-\omega } = S_\omega^\dagger ,
\la{eq:S}
\ee
where the set $\{\omega\}$ contains the ``Bohr frequencies'' of the Hamiltonian:
\be
H = \sum_k \epsilon_k | k \rangle \langle k |, ~~  \omega = \epsilon_k - \epsilon_l .  
\la{Bohr}
\ee
Then we can rewrite \Eq{eq:K(t)} as
\be
K(t) \rho = \sum_{\omega ,\omega'} S_\omega \rho S_{\omega'}^\dagger \int_0^t e^{i(\omega -\omega') u} du \int_{-u}^{t-u} F(\tau) e^{i\omega \tau} d\tau +(\rm{similar\ terms}) 
\la{eq:K2}
\ee
and use two crucial approximations:
\be
\int_0^t e^{i(\omega -\omega')u} du \approx t \delta _{\omega \omega'} ~~\hbox{and}~~
\int_{-u}^{t-u} F(\tau) e^{i\omega \tau} d\tau \approx G (\omega) = \int_{-\infty}^\infty F(\tau) e^{i\omega \tau} d\tau \geq 0 .
\la{eq:rep1}
\ee
This makes sense for $t \gg \max \{1/(\omega -\omega')\}$. Applying these two approximation we obtain $K(t) \rho_S = t \sum_\omega S_\omega \rho_S S_\omega^\dagger G(\omega) + \hbox{similar\ terms}$.  It follows from \Eq{eq:L} that $\cal L$ is a special case of the LGKS generator.  Returning to the Schr\"odinger picture one obtains the following Markovian master equation: 
\bea
\frac{d \rho}{dt} &=& - i [H,\rho] + {\cal L} \rho , \notag \\
{\cal L} \rho &\equiv& \frac{\lambda^2}{2} \sum_{\{\omega \}} G(\omega) \left( \left[ S_\omega , \rho S_\omega^\dagger \right] + \left[ S_\omega \rho ,S_\omega^\dagger \right] \right) .  
\la{eq:Dav}
\eea

Several remarks are in order:

\begin{enumerate}[label=(\roman*)]

\item The absence of off-diagonal terms in \Eq{eq:Dav}, compared to \Eq{eq:K2}, is the crucial property of the Davies generator, which can be interpreted as a coarse-graining in time of rapidly oscillating terms. It implies the commutation of $\cal L$ with the Hamiltonian part $[H, \cdot]$.

\item The positivity $G(\omega) \geq 0$ follows from Bochner's theorem and is a necessary condition for the complete positivity of the Markovian master equation.

\item For more complicated interaction Hamiltonians of the forms Eqs.\ (\ref{eq:hamint1}) and (\ref{eq:hamint2}), we must replace $G(\omega)$ by the positively defined relaxation matrix $G_{\alpha \beta}(\omega)$, which (because of symmetry) is usually diagonal in an appropriate parametrization.

\item The property (i) implies that the diagonal matrix elements of the density matrix (in the energy representation) evolve independently of the off-diagonal ones. They satisfy the Pauli Master Equation, with transition rates equal to those given by the Fermi Golden Rule. \cite{AL2007}

\end{enumerate}

If the reservoir is a quantum system in a thermal equilibrium state, the Kubo-Martin-Schwinger (KMS) condition holds:
\be
G(-\omega) = \exp \left( -\frac \omega {k_B T} \right) G(\omega) ,  
\la{eq:KMS1}
\ee
where $T$ is the bath's temperature.  As a consequence of \Eq{eq:KMS1}, the Gibbs state 
\be
\rho_\beta = Z^{-1} e^{-\beta H}, ~~\hbox{with}~~ \beta= \frac{1}{k_B T} ,
\la{eq:Gibbs}
\ee
is a stationary solution of \Eq{eq:Dav}. Under mild conditions (e.g., that the only system operators commuting with $H$ and $S$ are scalars) the Gibbs state is a unique stationary state and any initial state relaxes towards equilibrium (``zeroth law of thermodynamics'').  A convenient parametrization of the corresponding thermal generator reads
\be
{\cal L} \rho = \frac 1 2 \sum_{\{\omega\geq 0\} } \gamma(\omega) \left\{ \left( \left[ S_\omega, \rho S_\omega^\dagger \right] + \left[ S_\omega \rho, S_\omega^\dagger \right] \right) + e^{-\hbar \beta \omega} \left( \left[ S_\omega^\dagger, \rho S_\omega \right] + \left[ S_\omega^\dagger \rho, S_\omega \right] \right) \right\} , 
\la{eq:Dav_therm}
\ee
where finally 
\be
\gamma(\omega)= \lambda^2 \int_{-\infty}^{+\infty} \Tr \left( \rho_R \, e^{i H_R t/\hbar}\, R \, e^{-iH_R t/\hbar} R \right) \, dt .  
\la{eq:relaxation}
\ee
The generalization to more complicated forms of the interaction Hamiltonian is straightforward.

 \subsection{Spohn inequality and thermodynamic interpretation}
As the solutions of MME discussed above are given by one-parameter semigroups of {\it completely positive} maps, one can use the monotonicity of the relative entropy $S(\rho_1 | \rho_2) = \Tr \left[ \rho_1 \ln \rho_1 - \rho_1 \ln \rho_2 \right]$ in the form
\be
S(\Lambda\rho_1 | \Lambda \rho_2) \leq S(\rho_1 | \rho_2) ,
\la{eq:mono}
\ee
valid for any completely positive trace-preserving map $\Lambda$.  Identifying the von Neumann entropy $S(t) = - k_B \Tr [\rho(t)\ln\rho(t)]$ with the thermodynamic entropy and taking as $\rho(t)$ the solution of MME with a stationary state $\bar \rho$ one obtains from \Eq{eq:mono} the Spohn inequality
\be
\sigma(t) = - k_B \Tr \left[ \mathcal L \rho(t) \left( \ln \rho(t) - \ln \bar{\rho} \right) \right] \geq 0 .
\la{eq:spohn1}
\ee
This allows us to write the entropy balance, using a positive entropy production identified with $\sigma(t)$:
\be
\dot S(t) = \sigma(t) - k_B \Tr\left[\ln \bar \rho \frac{d}{dt}\rho(t)\right] . 
\la{eq:entprod}
\ee
If the Gibbs state of \Eq{eq:Gibbs} is stationary, then the second term on the RHS of \eqref{eq:entprod} can be written as $J/T$ where
\be
J = \frac{d}{dt}\Tr [\rho(t) H] .
\la{eq:heatflow}
\ee
This can be interpreted as a heat flow from the heat bath. Then \Eq{eq:entprod} becomes the standard form of the second law of thermodynamics for this particular open system.


\bibliographystyle{aipprocl}   

\end{document}